\documentclass[twocolumn,showpacs,amsmath,amssymb,prd,citeautoscript]{revtex4}

\usepackage{graphicx}
\usepackage{dcolumn}
\usepackage{bm}

\newcommand{\psib}{{\overline{\psi}}}
\newcommand{\fXY}{{\mathrm{fXY}}}
\newcommand{\cXY}{{\mathrm{bXY}}}

\begin{document}
\title{Absence of vortex condensation in a two dimensional fermionic XY model}
\author{D. J. Cecile and Shailesh Chandrasekharan}
\affiliation{Department of Physics, Box 90305, Duke University, Durham, 
North Carolina 27708.}
\date{\today}

\begin{abstract}
Motivated by a puzzle in the study of two dimensional lattice Quantum 
Electrodynamics with staggered fermions, we construct a two dimensional 
fermionic model with a global $U(1)$ symmetry. Our model can be mapped into a 
model of closed packed dimers and plaquettes. Although the model has the same 
symmetries as the $XY$ model, we show numerically that the model lacks the well
known Kosterlitz-Thouless phase transition. The model is always in the 
gapless phase showing the absence of a phase with vortex condensation. In other
words the low energy physics is described by a non-compact $U(1)$ field theory.
We show that by introducing an even number of layers one can introduce vortex 
condensation within the model and thus also induce a KT transition. 
\end{abstract}

\maketitle

\section{Motivation}

Two dimensional lattice Quantum Electrodynamics continues to be of interest today as a test bed for ideas and algorithms for lattice QCD \cite{Bietenholz:2007gb,Nagai:2007nj,Hasenfratz:2006qt,Christian:2005yp,Durr:2003xs}. In this work we focus on the formulation with staggered fermions, and refer to it as LQED2 \cite{Dilger:1995pi}. In the continuum limit, the theory is expected to describe the two-flavor Schwinger model \cite{Coleman:1976uz}. With massless fermions the two-flavor Schwinger model contains an $SU(2)\times SU(2)$ chiral symmetry. Away from the continuum limit, finite lattice spacing effects in LQED2 break the chiral symmetry to a $U(1)$ subgroup. In the mean field approximation this symmetry 
is spontaneously broken. However, since in two dimensions strong infrared 
fluctuations forbid spontaneous symmetry breaking, the mean field result is 
modified \cite{Mermin:1966fe,Coleman:1973ci}: Instead the theory develops 
critical long range (gapless) correlations, which can be detected through the 
chiral condensate susceptibility
\begin{equation}
\chi = \sum_i \Bigg\langle \psib_i\psi_i \ \psib_j \psi_j\Bigg\rangle 
\sim A L^{2-\eta}
\label{corr2}
\end{equation}
where $\psi_i$ and $\psib_i$ are the staggered fermion fields at the site $i$
 on a square lattice. In general one expects $\chi \sim A L^{2-\eta}$ in the
 gapless phase and $\chi \sim B$ when the theory develops a mass gap, where 
$A$ and $B$ are constants. In the case of a $U(1)$ symmetric theory, the mass 
gap can be generated only due to vortex condensation. In LQED2, at strong 
couplings one finds that Eq.~(\ref{corr2}) holds with $\eta=0.5$ 
\cite{Chandrasekharan:2003qv}. On the other hand in the continuum limit, 
using the result in the two-flavor Schwinger model, one expects $\eta=1$. 
Since previous studies have shown no evidence of a phase transition from 
strong to weak gauge couplings one might conjecture that the theory is always 
in the gapless phase with $\eta$ varying smoothly from $0.5$ to $1.0$ as a 
function of the gauge coupling. As far as we know this has not yet been shown
 analytically or observed numerically.

Although at first glance there does not seem to be anything strange with the 
above scenario, a closer examination reveals an interesting puzzle. Away from 
the continuum limit we expect the low energy physics of LQED2 to be described 
by a $U(1)$ symmetric bosonic field theory since the fermions are confined. Is
this a compact or a non-compact $U(1)$ field theory? A compact $U(1)$ theory 
will contain vortices, just like the $XY$ model, and these vortices can 
condense. Thus, it can undergo the famous Kosterlitz-Thouless(KT) phase 
transition, due to the presence of vortices, from a gapless phase to a massive 
phase. Further, although the susceptibility $\chi$ is governed by 
Eq.~(\ref{corr2}) in the gapless phase, $\eta$ satisfies the constraint 
$0 < \eta \leq 0.25$. Thus $0.5 \leq \eta \leq 1$, expected in LQED2, 
appears inconsistent with the physics of a compact $U(1)$ field theory. On the
 other hand it was shown recently that when one studies LQED2 at strong 
couplings in a slab geometry with four two-dimensional layers, one finds a
 KT transition separating a gapless phase and a gapped phase. Further in the 
gapless phase Eq.~(\ref{corr2}) holds and indeed one finds 
$0 < \eta \leq 0.25$ as expected \cite{Chandrasekharan:2003qv}. Thus, we
 conclude that LQED2 behaves like a non-compact $U(1)$ theory on a square 
lattice, but behaves like a compact $U(1)$ theory on a lattice where four 
two-dimensional layers are coupled to each other. {\em How is this possible?}
The motivation behind our work is to shed some light on this question by 
constructing a simpler model that can be easily studied numerically and 
that clearly exhibits all the above features.

Our paper is organized as follows: In Section \ref{fxy} we construct a new 
realization of the lattice XY model with fermionic composites. This model is 
easier to study than LQED2 but captures the essential physics. In Section 
\ref{plane} we show that our simpler model, on a square lattice, is always 
in the gapless phase and does not contain the Kosterlitz-Thouless (KT) phase
transition. In other words the condensation of vortices is eliminated 
completely suggesting the long distance theory is similar to non-compact 
$U(1)$ field theory. In Section \ref{slab} we show that adding one extra 
two dimensional layer leads to a gapped (massive) phase. This means the 
extra layer introduces vortex condensation. With four or more layers the 
theory contains both the gapless and the gapped phases separated by the 
usual KT transition. In Section \ref{concl} we present our conclusions and
suggest some directions for the future.

\section{A Fermionic XY Model}
\label{fxy}

The action of the conventional lattice XY ($\cXY$) model is given by
\begin{equation}
S = -\frac{1}{2} \sum_{<ij>} z_i^* z_j + z^*_j z_i
\label{cxymodel}
\end{equation}
$<ij>$ stands for the sum over nearest neighbor sites and 
$z_i = \exp(i\phi_i)$ is a complex variable of unit magnitude at site $i$ 
on a two-dimensional $L\times L$ square lattice with periodic boundary 
conditions.  The partition function of the model is given by
\begin{equation}
Z = \int [d\phi] \exp(- \frac{1}{T} S).
\label{cpf}
\end{equation}
Clearly the action and the measure of the partition function are invariant 
under the $U(1)$ symmetry $z \rightarrow \exp(i\theta) z$. The $\cXY$ model 
is known to contain vortices which leads to the existence of two phases as 
a function of $T$: a gapless phase where vortices are confined and a gapped 
phase where they are free and condense. The phase transition separating them 
is the well known KT transition and occurs at some critical temperature $T_c$. 
For $T < T_c$ for large $L$ one expects
\begin{equation}
\chi = \sum_i \Bigg\langle z_i z^*_j + z_j z^*_1 \Bigg\rangle \sim A L^{2-\eta}
\label{corr1}
\end{equation}
where $\eta$ depends on $T$. One of the important predictions of the KT 
transition is that $\eta \rightarrow 0.25$ as $(T_c-T) \rightarrow 0$.

Here we construct and study a new type of fermionic-$XY$ ($\fXY$) model 
constructed with Grassmann variables. The action of this model is given by
\begin{equation}
S_1 = -\sum_{<ij>} \psib_i\psi_i \psib_j\psi_j 
- \beta \sum_{<ijkl>} \psib_i\psi_i \psib_j\psi_j \psib_k\psi_k \psib_l\psi_l
\label{fxymodel}
\end{equation}
where $\psi_i$ and $\psib_i$ are two independent Grassmann variables on the 
site $i$ and $<ijkl>$ stands for all plaquettes made up of sites $i,j,k,l$ in
a cyclic order. The partition function is given by
\begin{equation}
Z = \int [d\psi d\psib] \exp(- S_1).
\label{fpf}
\end{equation}
Note that an overall multiplicative constant $1/T$, like in Eq.(\ref{cpf}), 
drops out of all observables up to multiplicative constants, if $\beta$ is 
redefined as $T\beta$. The action and the measure in eq.~(\ref{fpf}) are again
 invariant under the $U(1)$ symmetry $\psi \rightarrow \exp(i\sigma_i \theta)
 \psi$ and $\psib \rightarrow \exp(i\sigma_i \theta) \psib$ where $\sigma_i =
 1$ on all even sites and $-1$ on all odd sites. Here we ask if the low energy 
theory of the $\fXY$-model resembles that of the $\cXY$-model. In particular, 
as a function of $\beta$, are there two phases separated by the KT phase 
transition?

\begin{figure}[t]
\begin{center}
\includegraphics[width=0.47\textwidth]{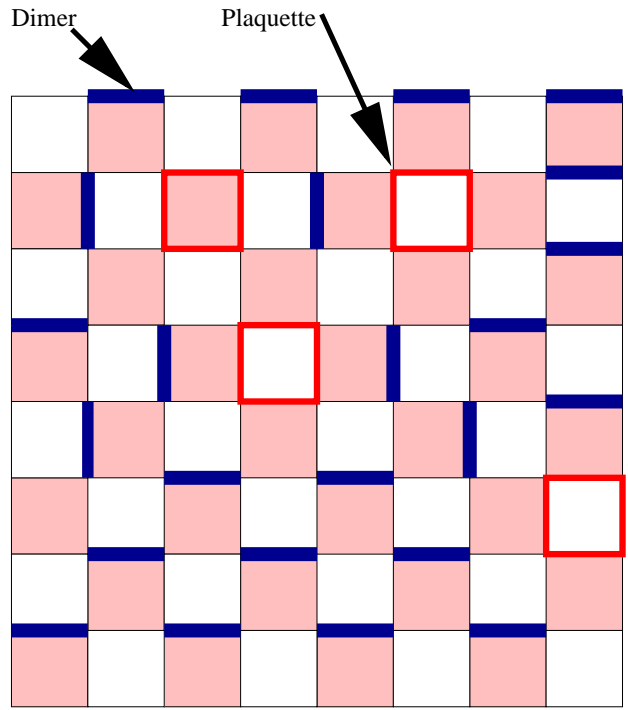}
\end{center}
\caption{\label{fig1} A example of a dimer-plaquette configuration of the
fermionic $XY$ model.}
\end{figure}

It is possible to integrate out the Grassmann variables and rewrite the 
$\fXY$-model as a statistical mechanics of bond variables (referred to as 
dimers) $b$ and plaquette variables $p$. In this representation, the partition
 function is given by
\begin{equation}
Z = \sum_{[b,p]} \prod_{<ijkl>} \beta^{p_{<ijkl>}}
\end{equation} 
where $b_{<ij>}=0,1$ and $p_{<ijkl>}=0,1$ are the allowed values with the 
constraint that only one of the four $b$'s or $p$'s associated with a given 
lattice site must be non-zero. Physically, this constraint means that each 
site be connected to only one dimer or one plaquette. An example of the 
dimer-plaquette configuration is shown in Fig.\ref{fig1}.

When $\beta=0$ the $\fXY$-model defined above is exactly solvable and was 
first studied in \cite{Fisher}. It was found that the chiral condensate 
susceptibility
\begin{equation}
\chi = \sum_i \Bigg\langle \psib_i\psi_i \psib_j \psi_j\Bigg\rangle
\label{corr3}
\end{equation}
satisfies eq.~(\ref{corr2}) with $\eta=0.5$. Since $\eta$ does not lie 
between $0$ and $0.25$, the $\fXY$ model is clearly different from the $cXY$ 
model. Below we will argue this difference in behavior results from 
elimination of vortex condensation completely.

\section{Absence of vortex condensation}
\label{plane}

In order to study the $\fXY$-model we have developed a directed path algorithm which is a straight forward extension of the ideas presented in \cite{Adams:2003cc}. Here we focus on the results and postpone the discussion of the algorithm to another publication. One of the features of the algorithm is that it allows us to measure $\chi$ efficiently. In Tab.(\ref{tab1}) we compare the results of the algorithm with exact calculations on small lattices for various values of $\beta$.

\begin{table}[ht]
\begin{center}
\begin{tabular}{|c|c|c|c|} \hline
Lattice size & $\beta$ & Exact      & Algorithm \\\hline
$6\times6$   &  0.0    & 3.33640... & 3.3366(9) \\ 
$4\times4$   &  0.3    & 1.43103... & 1.4307(4) \\
$4\times4$   &  5.5    & 0.34615... & 0.3461(2) \\
$6\times6$   &  10.0   & 0.30663... & 0.3068(5) \\
$6\times6$   &  100.0  & 0.02888... & 0.0288(2) \\\hline
\end{tabular}
\caption{\label{tab1} Comparison of $\chi$ obtained using the directed path 
algorithm with exact results. The algorithm appears to remain efficient for 
relatively large values of $\beta$ on small lattices.}
\end{center}
\end{table}

\begin{figure}[t]
\begin{center}
\includegraphics[width=0.47\textwidth]{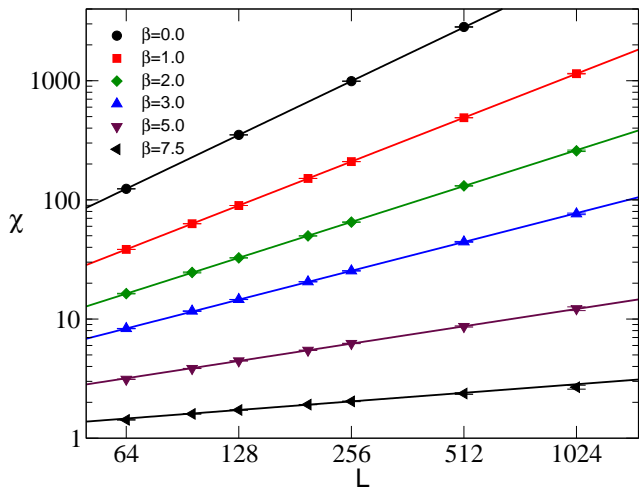}
\end{center}
\caption{\label{fig2} Chiral susceptibility as a function of $L$ for a variety
 of $\beta \leq 7.5$. For larger $\beta$ we find it difficult to reach the 
thermodynamic limit even though the algorithm may be efficient.}
\end{figure}

In Fig.(\ref{fig1}) we plot $\chi$ as a function of $L$ for various values 
of $\beta$ ranging from $0$ to $7.5$. We believe the algorithm remains stable for these values of $\beta$. We find that the behavior of $\chi$ as a function of $L$ fits well to the form $A L^{2-\eta}$, expected in the gapless phase, for all values of $\beta$. The values of $A$ and $\eta$ obtained from the fits are tabulated in Tab.(\ref{tab2}).
\begin{table}[ht]
\begin{center}
\begin{tabular}{|l|l|l|r|} \hline
\em $\beta$ &\em $A$ &\em $\eta$ & $\chi^2$/DOF \\\hline
0.0 & 0.242(1) & 0.5      & 1.3 \\
1.0 & 0.236(1) & 0.776(1) & 0.5 \\
2.0 & 0.256(3) & 1.000(2) & 0.7 \\
3.0 & 0.294(3) & 1.196(2) & 1.3 \\
5.0 & 0.428(8) & 1.517(4) & 0.3 \\
7.5 & 0.54(1)  & 1.765(5) & 1.3 \\
\hline
\end{tabular}
\caption{\label{tab2} Values of $A$ and $\eta$ obtained by fitting $\chi$ to 
the form $A L^{2-\eta}$.}
\end{center}
\end{table}
As Fig.(\ref{fig2}) and the fits in Tab.(\ref{tab2}) clearly show, there is 
no sign of a phase transition as a function of $\beta$ on lattices up to 
$L=1024$. The value of $\eta$ slowly rises with $\beta$. Based on locality of
the theory we expect that in the $\beta \rightarrow \infty$ limit we should 
get $\eta = 2$. Thus, there is no gapped phase in the $\fXY$ model as a 
function of $\beta$: Vortices do not condense and the model behaves like a 
non-compact $U(1)$ field theory.

\section{Introducing vortices}
\label{slab}

How can we introduce vortex condensation in the $\fXY$ model? This is indeed 
possible if we consider $N>1$ two-dimensional layers of square lattices, 
coupled through a local interaction. The action of the $N$-layered model is 
given by
\begin{equation}
S_N = S_1 - t \sum_{i,<ll'>} \psib_{i,l}\psi_{i,l} \psib_{i,l'}\psi_{i,l'},
\end{equation}
where, in addition to the action given in Eq.~(\ref{fxymodel}), we have added
a term that couples neighboring layers $l$ and $l'$ represented by $<ll'>$ at
each site $i$. In the dimer-plaquette language the extra term introduces 
additional dimers that connect two neighboring layers at a site. We will refer
to the fermionic model with $N$ layers as $\fXY_N$. Thus the model 
$\fXY_1 \equiv \fXY$ is the model we considered above which does not contain 
vortex condensation. Here we will show that $\fXY_2$ and $\fXY_4$ contain a 
gapped phase unlike the $\fXY_1$ model. Here we focus at $\beta=0$ and study
the models as a function of $t$ and $N$.

\begin{figure}[t]
\begin{center}
\includegraphics[width=0.47\textwidth]{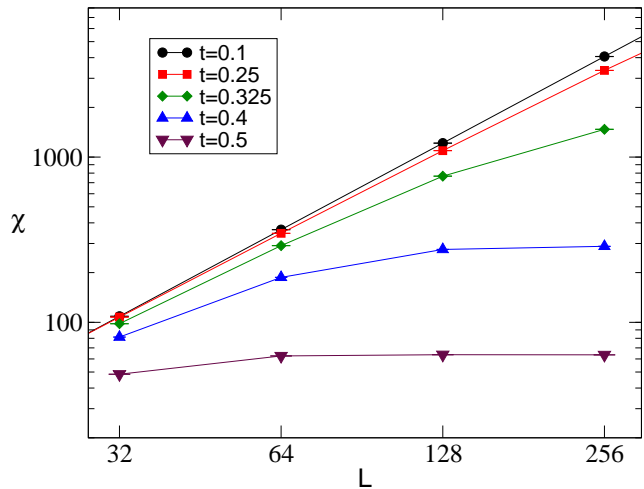}
\end{center}
\caption{\label{fig3} $\chi$ as a function of $L$ for $N=2$. The saturation 
of $\chi$ for large $t$ implies the presence of a finite correlation length. 
The fits are consistent with a diverging correlation length as 
$t\rightarrow 0$.}
\end{figure}

Let us first consider the model with two layers ($N=2$). Figure \ref{fig3} 
shows $\chi$ as a function of $L$ for various values of $t$. Table \ref{tab3}
 gives the value of $A$ and $\eta$ obtained from a fit of the data to the 
form $A L^{2-\eta}$.
\begin{table}[h]
\begin{center}
\begin{tabular}{|l|l|l|r|} \hline
\em $t$ &\em $A$ &\em $\eta$ & $\chi^2$/DOF \\\hline
0.01 & 0.185(1) & 0.252(1) & 2.2 \\
0.1 & 0.261(1) & 0.259(1) & 0.4 \\
0.25 & 0.337(1) & 0.335(1) & 419 \\
\hline
\end{tabular}
\caption{\label{tab3} Values of $A$ and $\eta$ obtained from fitting the
 behavior of $\chi$ as a function of $L$, in the range $32 \leq L \leq 256$,
 to the form $A L^{2-\eta}$ for different values of $t$ but $N=2$. The fit 
fails for $t \geq 0.25$.}
\end{center}
\end{table}
The fits begin to fail for $t \geq 0.25$, while they are good for $t\leq 0.1$.
The saturation of $\chi$ for large $L$ at $t=0.5$ is consistent with the
presence of a finite correlation length which implies the existence of a 
gapped phase due to vortex condensation. As $t$ decreases the 
correlation length increases. Is there a phase transition at a finite $t$? 
Note that the value of $\eta$ approaches $0.25$ as $t$ is lowered to zero. 
Interestingly, as discussed earlier, $\eta=0.25$ is a universal result at the 
critical point in a KT phase transition. Thus, it is likely that the 
correlation length diverges only at $t\rightarrow 0$ and not at any finite 
$t$. The reason for a good fit for $t \leq 0.1$ could be due to the fact 
that the correlation lengths at these values of $t$ are much larger than the 
lattice sizes explored. 

\begin{figure}[t]
\begin{center}
\includegraphics[width=0.47\textwidth]{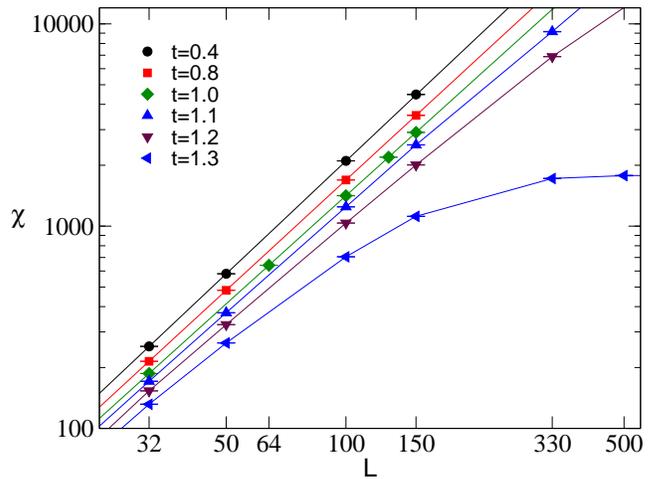}
\end{center}
\caption{\label{fig4} Chiral susceptibility as a function of $L$ for $N=4$. 
The solid lines for $t=0.4$, $0.8$ and $1.0$ are fits to $A L^{2-\eta}$ given 
in the text. This power-law fit begins to fails for $t=1.1$ and above. In 
particular, a KT transition is expected at $t=t_c\sim 1.05$ 
\cite{Chandrasekharan:2003qv}.}
\end{figure}

Next we consider $N=4$, which was studied earlier in 
\cite{Chandrasekharan:2003qv}. In Fig.( \ref{fig4}) we plot $\chi$ as a 
function of $L$ for various values of $t$. Table \ref{tab4} contains the
values of $A$ and $\eta$ obtained from fits to the data as in the $N=2$ case.
\begin{table}[ht]
\begin{center}
\begin{tabular}{|l|l|l|r|} \hline
\em $t$ &\em $A$ &\em $\eta$ & $\chi^2$/DOF \\\hline
0.4 & 0.412(2) & 0.146(1) & 1.3  \\
0.8 & 0.402(2) & 0.188(1) & 0.1  \\
1.0 & 0.396(1) & 0.224(1) & 1.1  \\
1.1 & 0.410(1) & 0.259(1) & 2.8  \\
1.2 & 0.487(1) & 0.338(1) & 124  \\
1.3 & 1.004(5) & 0.586(1) & 1500 \\
\hline
\end{tabular}
\caption{\label{tab4} Values of fitting coefficients $A$ and $\eta$ obtained 
by fitting $\chi = A L^{2-\eta}$ for different values of $t$. The range 
$32 \leq L \leq 150$ was used in the fit.}
\end{center}
\end{table}
The evidence from the fits is consistent with the following scenario: For 
$t\leq 1.0$ the model is in a gapless phase (absence of vortex condensation),
while for $t>1.1$ the model is in a gapped phase (existence of vortex 
condensation). The value of $\eta$ close to the critical point, 
$t_c \sim 1.05(5)$, is again about $0.25$, consistent with a KT phase 
transition. Thus, for $N=4$ the model contains a KT phase transition.
For even $N > 4$ we have evidence, from other unpublished studies, that the 
behavior is similar to that of $N=4$.

\section{Conclusions}
\label{concl}

In this work we have constructed and studied a fermionic $XY$ model. When the 
model is studied on a square lattice it contains no evidence for a gapped phase
 with vortex condensation. For this reason we believe it gives a realization 
of a two dimensional non-compact $U(1)$ field theory. Introducing an even 
number of layers, leads to a gapped phase phase which must be accompanied by 
vortex condensation. LQED2, away from the continuum limit, gives another 
realization of the long distance physics of our fermionic $XY$ model physics. 
Our work sheds light on why the low energy physics of LQED2 cannot be described
 by the $\cXY$ model even though the symmetries are the same.

\begin{figure}
\begin{center}
\includegraphics[width=0.47\textwidth]{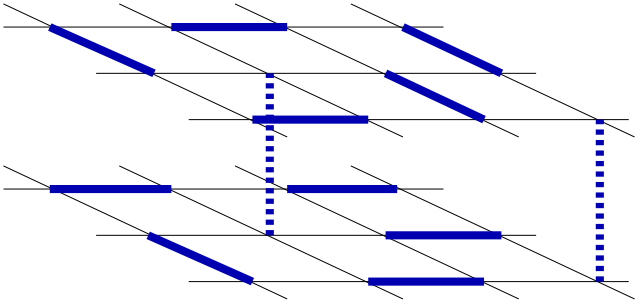}
\end{center}
\caption{\label{fig5} A dimer confinguration in two layers. The sites with 
{\em dashed} dimers connecting the layers is conjectured to be the location 
of the vortex. Note that for all $t>0$ the density of such sites will be 
non-zero leading to a condensation of vortices \cite{Chandrasekharan:2003qv}.}
\end{figure}

There are several questions that may be of further interest. For example:
\begin{enumerate}
\item Is it possible to characterize a vortex in the dimer-plaquette 
configuration? We conjecture that a vortex core may be identified by the
existence of a site such that, on that site all the layers are connected by 
dimers to the neighboring layers. An example of such a site is shown in 
Fig.(\ref{fig5}) for two layers. 
\item Is there some difference about even and odd values of $N$ since our 
definition of the vortex core does not work if $N$ is odd. In one layer we 
have already seen lack of vortex condensation. So what happens in $N=3,5,...$?
\item Does the two layered model approach a $KT$ critical point as 
$t\rightarrow 0$ as conjectured here? Can this result be understood through 
analytic means? 
\item What is the effect of $\beta > 0$ in the layered model. Hopefully, 
this will only introduce more disorder in the system. This may be worth 
studying further since a new type of order may be established for large 
$\beta$. 
\end{enumerate}
The answers to these questions could shed more light on the field theories described by the dimer-plaquette models.

\section*{Acknowledgements}
We thank Harold Baranger and Uwe-Jens Wiese for helpful discussions. SC thanks 
Professor Tao Pang at the University of Nevada in Las Vegas, for hospitality 
when this work was being completed. This work was supported in part by the 
Department of Energy grant DE-FG02-05ER41368 and by the National Science
Foundation grant DMR-0506953.

\bibliography{dimer}

\end{document}